\documentclass[
    reprint,
    superscriptaddress,
    amsmath,
    amssymb,
    aps,
    prx,
    nofootinbib,
    longbibliography,
]{revtex4-2}
\usepackage{silence}
\ActivateWarningFilters[revtex4-2]

\usepackage{graphicx}
\usepackage[dvipsnames]{xcolor}
\usepackage[USenglish]{babel}
\usepackage{mathtools}
\usepackage[colorlinks]{hyperref}
\usepackage[capitalize]{cleveref}
\usepackage{mleftright}
\usepackage[per-mode=reciprocal]{siunitx}
\usepackage[shortlabels]{enumitem}

\usepackage{macros}

\hypersetup{%
    colorlinks=true,
    linkcolor=Blue,
    linkbordercolor=Blue,
    citecolor=Blue,
    citebordercolor=Blue,
    filecolor=BrickRed,
    filebordercolor=BrickRed,
    urlcolor=NavyBlue,
    urlbordercolor=NavyBlue,
    menucolor=BrickRed,
    menubordercolor=BrickRed,
    runcolor=BrickRed,
    runbordercolor=BrickRed,
}

\begin{document}

\title{Chiral cavities made from lattices of highly electromagnetically-chiral scatterers}

\author{Lukas Rebholz}
\affiliation{%
    Institute of Theoretical Solid State Physics,
    Karlsruhe Institute of Technology,
    Kaiserstr. 12, 76131 Karlsruhe, Germany
}
\author{Carsten Rockstuhl}
\affiliation{%
    Institute of Theoretical Solid State Physics,
    Karlsruhe Institute of Technology,
    Kaiserstr. 12, 76131 Karlsruhe, Germany
}
\affiliation{%
    Institute of Nanotechnology,
    Karlsruhe Institute of Technology,
    Kaiserstr. 12, 76131 Karlsruhe, Germany
}
\author{Ivan Fernandez-Corbaton}
\affiliation{%
    Institute of Nanotechnology,
    Karlsruhe Institute of Technology,
    Kaiserstr. 12, 76131 Karlsruhe, Germany
}

\date{July 14, 2025}

\begin{abstract}
The infamous weakness of molecular chiroptical responses challenges the all-optical realization of crucial applications such as enantio-selective sorting of chiral molecules, or biasing chiral chemical reactions.
Chiral optical cavities are a natural choice for confronting this challenge.
Ideally, the dissymmetry between the two helicities inside such cavities is maximized.
In here, we propose a chiral infrared optical cavity formed by planar mirrors made of diffracting lattices of silver helices with almost maximum electromagnetic chirality.
It combines the strong helicity selectivity of the helices with the helicity-preserving reflectivity that planar systems show at large incidence angles.
For the manifold of cavity modes which have a component with zero in-plane momentum, we demonstrate an unprecedented dissymmetry of \qty{95}{\percent} inside the cavity at the target frequency, making it a compelling candidate for enantio-selective applications.
\end{abstract}

% Use showkeys class option if keyword display desired

\maketitle

The ubiquitous presence of chirality across physical scales and its importance in chemical and biological processes make the understanding and control of chirality a major endeavor across scientific disciplines. 
The best understood chiral entity is the electromagnetic helicity, or polarization handedness of the electromagnetic field, which has been used for centuries to probe the chirality of matter. 
By leveraging this understanding in combination with modern fabrication techniques, chiral optical cavities \cite{Taradin2021,Huebener2021,Voronin2022,Gautier2022}, which can resonantly enhance chiral light--matter interactions, are poised to become indispensable in our quest to control the chirality of matter.

The chiral environment created inside a chiral cavity enables its enantio-selective interaction with other chiral entities, such as the two helicities of the electromagnetic radiation of an emitter, or the two enantiomers of chiral molecules. 
One can utilize chiral modes to create circularly polarized lasers \cite{Maksimov2022,katsantonis_ultrathin_2025} or enhance the chirality of luminescent emissions \cite{Seongheon2023}.
Separating the two enantiomers of a chiral molecule by optical means is another active research area \cite{Canaguier2013,Solomon2019,MartinezRomeu2024}. 
While such chiral sorting is possible for larger objects \cite{Tkachenko2014b}, the infamous weakness of molecular chirality poses a major challenge, and chiral cavities are one option for confronting it. 
Finally, the modification of ground-state chemistry inside cavities \cite{Hutchison2012,Thomas2016,Sau2021} suggests the use of chiral cavities for biasing chemical reactions towards one of the two enantiomers of the product \cite{Riso2024}. 
While there is currently no theory that fully explains such chemistry, it is reasonable to think that, for this and all the other enantio-selective applications, one should aim to maximize the chirality of the environment inside the cavity. 
This is the objective of this Letter.

In here, we consider an optical cavity to maximize the differential interaction between the cavity modes and the two enantiomers of given chiral molecules at a particular frequency.
To choose the frequency, we note that the absorption resonances of organic chiral molecules are grouped into three frequency bands: electronic transitions in the UV-visible region, vibrational lines in the ``fingerprint'' infrared region, and rotational lines in the microwave region. 
As the frequencies decrease by two orders of magnitude between bands, so does the relative differential absorption of the two handedness of light by the molecules, which is often quantified by the dimensionless Kuhn dissymmetry factor $g$: Typical values for $g$ are on the order of \num{e-2}, \num{e-4}, and \num{e-6} for the electronic, vibrational, and rotational lines, respectively \cite{Mason1963,Salzman1991}. 
In cavities tuned to the electronic transitions, the distance between the mirrors is less than \SI{1}{\micro\meter}, which severely compromises the flow of liquid in and out of the cavity. 
This leaves the infrared band as the next best option, which we pursue. 

The cavity will be formed by two planar mirrors facing each other. 
Ideally, all the modes of the cavity should be of the same pure single handedness \cite{Taradin2021,Huebener2021}.
While such an ideal cavity seems out of reach, we nevertheless can use the concept to guide our design. 
Accordingly, the mirrors should, in a helicity-preserving manner \cite{yan_3d-printed_2025}, reflect only one of the two helicities of light \cite{Plum2015,Semnani2020}.
This can be approached when the mirrors are diffracting lattices of scatterers with maximal electromagnetic chirality (em-chirality) \cite{fernandez-corbaton_objects_2016}. 
First, the lattice spacing is to be selected so that the first diffraction orders at the target frequency feature a large in-plane wave vector component. 
Large in-plane momentum leads to helicity-preserving reflections in many cases, for example, at the interface between any two homogeneous media \cite[App.~A]{feis_helicity-preserving_2020}, and also at more complex interfaces, including anisotropic media and (diffracting) gratings \cite[Chap. 2.3]{Lekner2016}\cite{Nakayawa2011,feis_helicity-preserving_2020}. 
Second, a scatterer that saturates its em-chirality, which is bounded by the square root of its total interaction cross-section \cite{fernandez-corbaton_objects_2016}, interacts exclusively with light of one helicity while being transparent to the other. 
The concept of em-chirality solves many of the problems that arise in chirality quantification \cite{Fowler2005,Vavilin2022}, and becomes a cost function for optimization \cite{Gorkunov2020,Arens2021,Kuehner2023}.
Optimized metallic helices are being studied and fabricated with the aim of maximizing their chiroptical response \cite{Kuen2025,Tsarapkin2025}, and have been numerically shown \cite{garcia-santiago_maximally_2022} to achieve large values of em-chirality for target frequencies below \SI{100}{\tera\hertz}, covering the molecular ``fingerprint'' region.
In the following, we first present the optimized metallic helix and then study the mirrors used for the cavity. 
Specifically, we analyze the helicity-preserving reflectivity of the mirrors and explain the physical origin of the observed characteristics. 
Finally, we present the chiral cavity formed by two of these mirrors and evaluate the helicity content of its eigenmodes, focusing on the manifold of eigenmodes that have a component with zero in-plane momentum. 
For this manifold, at the target frequency of $\qty{44.21}{\THz}$, and when the mirrors are separated by $\qty{16.23}{\micro\meter}$, the dissymmetry of the modal landscape inside the cavity reaches \qty{95.3}{\percent}. 
This unprecedented imbalance makes this cavity a compelling candidate for enantio-selective applications.
Such applications include biasing chiral reactions, where the current lack of theory leaves only the experimental route, which one would like to take with the most chiral cavity.

We start by identifying the geometrical parameters of a silver helix that maximize its em-chirality. 
We consider a helix embedded in a homogeneous, isotropic, and nonabsorbing ambient medium with a relative permittivity of $\permittivity_{\text{amb}} = \num{1.8912}$.
The ambient medium is chosen to account for the environment within the cavity, which is defined by substrates and a solution-filled cavity interior.
The material parameters for silver are taken from Ref.~\onlinecite{hagemann_optical_1975}\footnote{Data provided at \url{https://refractiveindex.info}. \cite{polyanskiy_refractiveindexinfo_2024}}.
In here, the em-chirality is formulated in a monochromatic picture in terms of the object's transition matrix, or T-matrix \cite{waterman_matrix_1965,waterman_symmetry_1971,mishchenko_peter_2013}, which fully represents an object's linear optical response at a fixed frequency. 
The target frequency is \qty{44.21}{\THz}, which is a resonance of BINOL \cite{scott_enhanced_2020} in the infrared region.
To identify the optimal geometrical parameters, we use Bayesian optimization with Gaussian processes as implemented in the Analysis and Optimization Toolkit provided by the simulation software \pkg{JCMsuite}\footnote{\pkg{JCMsuite} version 6.0.10, \href{https://jcmwave.com/}{JCMwave GmbH}.}.
The process involves computing the T-matrix from \ac{FEM} scattering simulations and subsequently evaluating the em-chirality.
More details on the numeric procedure to obtain T-matrices as well as the optimization are given in \cref{sec:tmatrix_fem,sec:optimization}.

\begin{figure}
    \centering
    \includegraphics[width=1.0\linewidth]{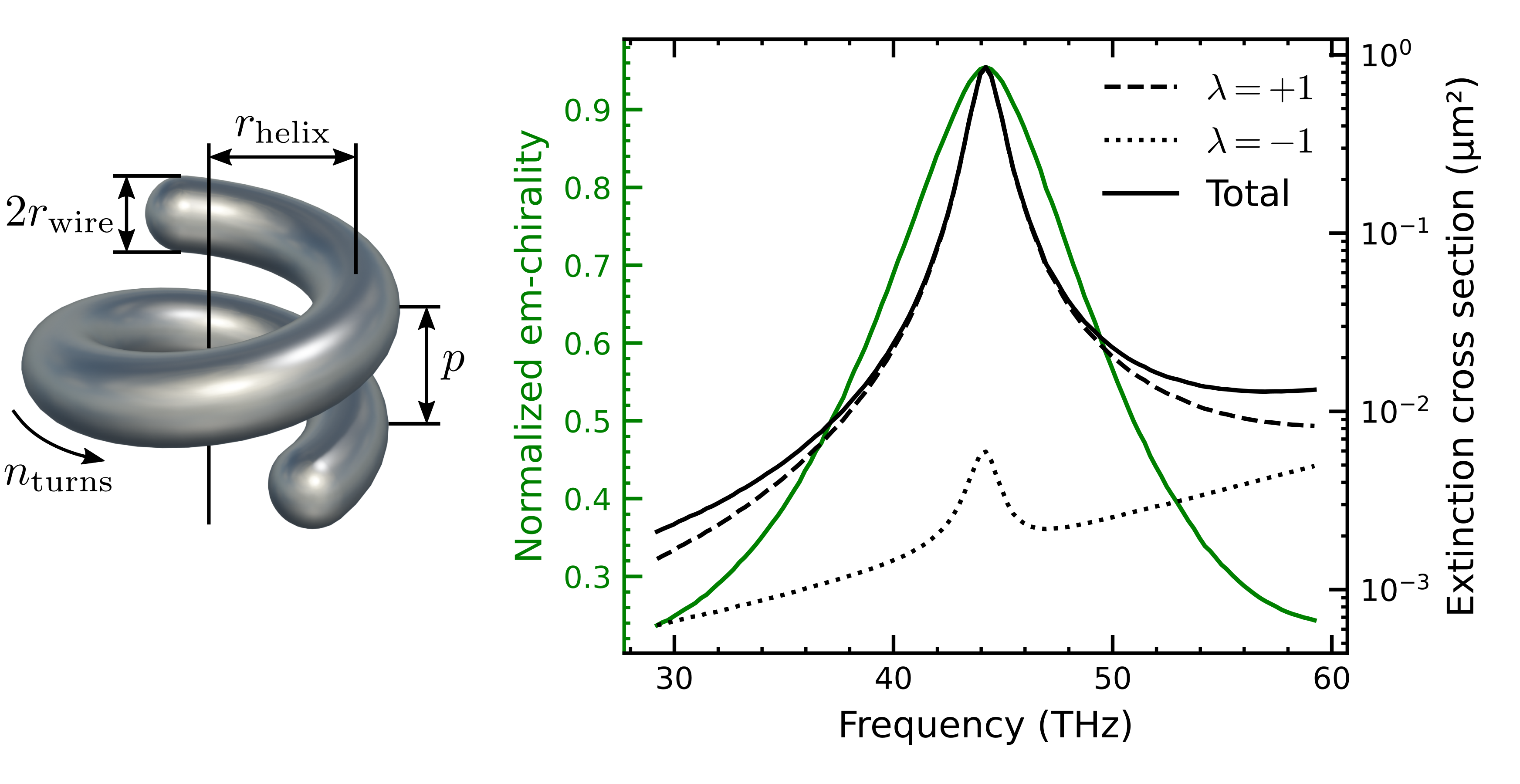}
    \caption{Silver helix with high em-chirality. 
    The green solid line shows the normalized em-chirality of a silver helix with parameters ${r_{\text{helix}} \approx \qty{183.0}{\nm}}$, ${r_{\text{wire}} \approx \qty{53.3}{\nm}}$, pitch ${p \approx \qty{168.6}{\nm}}$, and ${n_{\text{turns}} \approx \num{1.407}}$. 
    At the target frequency of $\qty{44.21}{\THz}$, the normalized em-chirality reaches its maximum of approximately \num{0.955}, so \qty{4.5}{\percent} shy of the absolute bound where the object is transparent to one of the two helicities of light. 
    The rotationally averaged extinction cross-section (solid black line) shows that the high em-chirality arises from a resonance in the helix's response. 
    The contrast in the extinction cross-section for incident light of positive helicity (dashed black) and negative helicity (dotted black) is two orders of magnitude.}
    \label{fig:em-chiral-helix}
\end{figure}

We obtain a helix that reaches an em-chirality, normalized to its total interaction cross-section, of approximately \num{0.955} at the target frequency. 
The optimal geometric parameters are given in Fig.~\ref{fig:em-chiral-helix}, which shows the spectrally resolved normalized em-chirality.
It shows the em-chirality to be a spectrally broad feature, indicating that the proposed cavity's ability to sustain a large helicity preference is not going to be limited to a narrow spectral range. 
\Cref{fig:em-chiral-helix} additionally shows the rotationally averaged extinction cross-section, unveiling the large em-chirality to be a resonant feature of the helix. 
At the point of maximum em-chirality, the helix exhibits a contrast in the helicity-dependent extinction cross-section reaching two orders of magnitude.
The response of the helix can be effectively understood as follows: Along its axis, the helix acts as an electric dipole antenna with charges moving from tip to tip.
Simultaneously, it forms an electric coil, giving rise to a magnetic dipole as charges turn along the helical path. 
The phase relation between the electric and magnetic dipole is determined by the handedness of the helix. 
If the geometry of the helix is properly adjusted, which in this work is achieved via optimization, its multipolar response becomes practically that of a helical dipole, whose electric and magnetic dipole moments, $\vec{d}$ and $\vec{m}$ respectively, are related as $\vec{m} = \pm\ii\SpeedOfLight\vec{d}$. 
The radiation from a helical dipole is of pure helicity.

\begin{figure}
    \centering
    \includegraphics[width=1.0\linewidth]{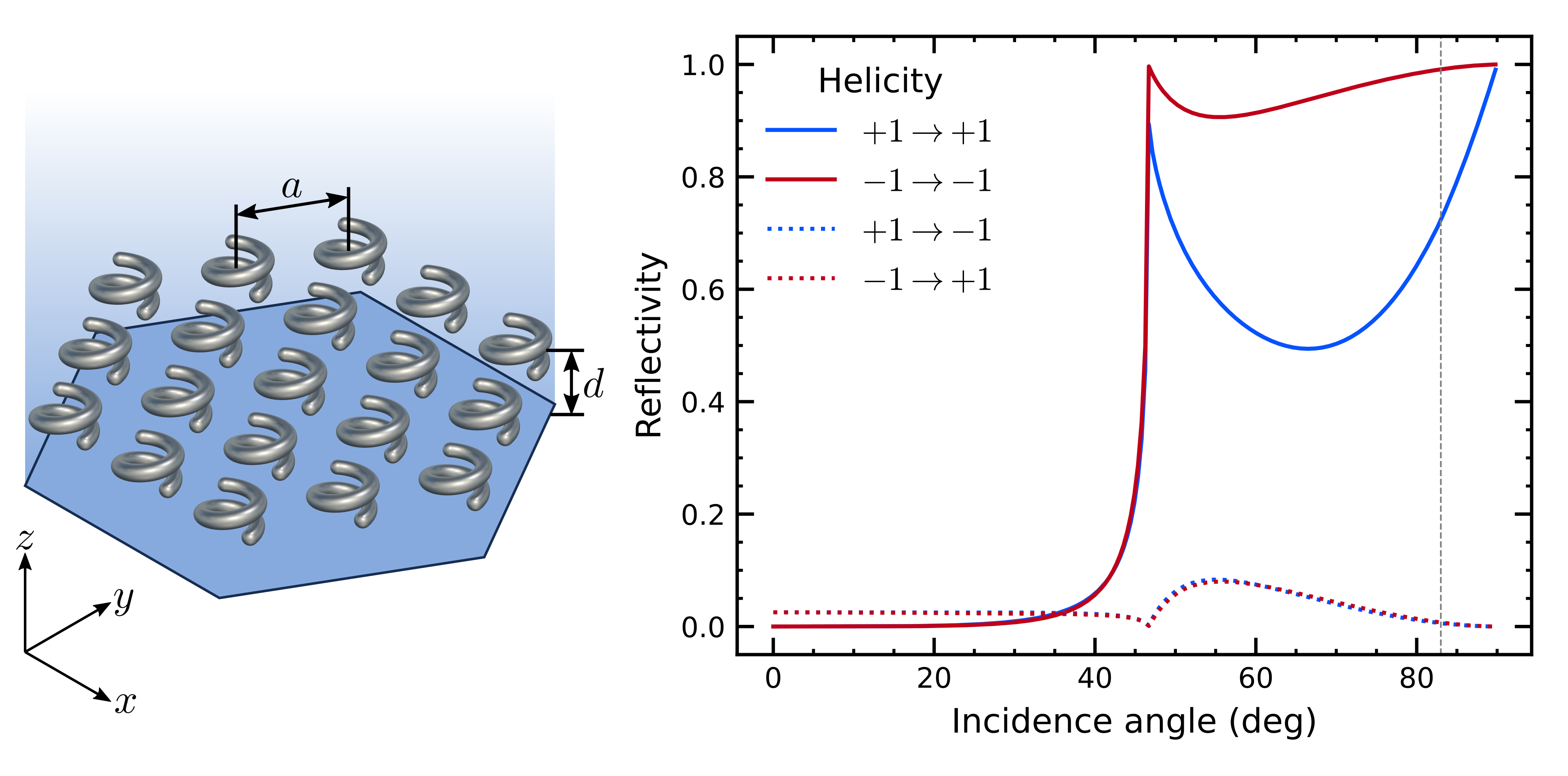}
    \caption{Reflectivity of the chiral mirror at the target frequency of \qty{44.21}{\THz}. 
    The mirror consist of a hexagonal lattice of em-chiral helices with a lattice spacing of $a \approx \qty{5.74}{\um}$ (size of helix and lattice spacing not shown to scale), situated at a distance of $d \approx \qty{1.12}{\um}$ above the interface between embedding and vacuum.
    Blue lines correspond to incident light of positive helicity, red lines to negative helicity.
    Solid lines show the helicity-preserving specular (zeroth-order) reflectivity, and dotted lines show the helicity-flipping reflectivity, which is small in the entire angle range.
    The gray dashed vertical line marks an incidence angle of \ang{83}.}
    \label{fig:chiral-mirror}
\end{figure}

We now move on to describe the mirrors from which the cavity is formed.
A single mirror for the proposed cavity consists of a hexagonal lattice of the em-chiral helix on top of a substrate matching $\permittivity_{\text{amb}} = 1.8912$.
A schematic depiction is shown in \cref{fig:chiral-mirror}.
To consider prospective applications of the cavity with the cavity interior filled, we assume the half-space above the mirror to be filled with some material.
For simplicity, we consider the case where the material is index-matched to the mirror substrates at $\permittivity_{\text{amb}} = 1.8912$.
However, index matching is not essential, provided that the refractive index of the cavity interior does not exceed that of the surrounding substrates; otherwise, \ac{TIR} will occur at the interface, affecting our analysis of how the cavity works.
The lattice spacing is chosen at $a \approx \qty{5.74}{\um}$.
At \qty{44.21}{\THz}, this puts the six first diffraction orders, relative to normal incidence, at an angle of \ang{83} to the surface normal, determining the sought-after large in-plane momentum component.
 
Importantly, the lattice is only diffracting in the embedding; relative to the vacuum wavelength, the lattice spacing is sub-wavelength.
The distance between the lattice of helices and the lower vacuum--embedding interface of the substrate, accounting for the thickness of the substrate plus half the height of a helix, is chosen at approximately \qty{1.12}{\um}. 
This value optimizes helicity-preservation in the reflection of the first diffraction orders.

In \cref{fig:chiral-mirror}, we study the specular reflectivity, in particular, the helicity-preserving reflectivity, of the cavity mirror at the target frequency in dependence of the incidence angle.
The considered incident angular wave vectors lie in the $k_{x}k_{z}$-plane with $k_{y} = 0$, with the mirror being oriented perpendicular to the $z$-axis and one of the lattice vectors $\parallel \unitvec{x}$.
While the mirror is not rotationally symmetric around the $z$-axis, as the helix breaks the lattice's sixfold rotational symmetry, the reflectivity characteristic shown in \cref{fig:chiral-mirror} does not significantly depend on the in-plane orientation of the illumination and is indicative for an illumination under any azimuthal angle.
The reflectivity at large incidence angles is of particular interest, as it is essential for sustaining modes with large in-plane momentum, or wave vector component.
The cavity mirror exhibits high reflectivity at large incidence angles, aided by \ac{TIR} at the substrate--vacuum (or air) interface.

\Ac{TIR} occurs at $\arcsin\mleft(\frac{1}{\sqrt{\permittivity_{\text{amb}}}}\mright) \approx \ang{46.6}$, visible as a steep rise in the reflectivity in \cref{fig:chiral-mirror}.
Crucially, the mirror exhibits a strong dissimilarity in its response to the different helicities due to the helix lattice.
The lattice introduces a highly helicity-selective loss channel through absorption and diffraction. 
The latter allows light to bypass \ac{TIR}, predominantly for light of positive helicity.
This lowers the helicity-preserving reflectivity for positive helicity at angles permitting \ac{TIR} to values as low as \qty{50}{\percent}, whereas the negative helicity counterpart stays well above \qty{90}{\percent}.
This imbalance allows the cavity modes with large transverse momentum to have a preferred helicity.
For example, when the incident angle is \ang{83}, the negative helicity experiences an almost perfect helicity-preserving reflectivity of \qty{99.2}{\percent}, while the positive helicity is attenuated by a factor of \qty{72.6}{\percent} at each reflection.

\begin{figure}
    \centering
    \includegraphics[width=1.0\linewidth]{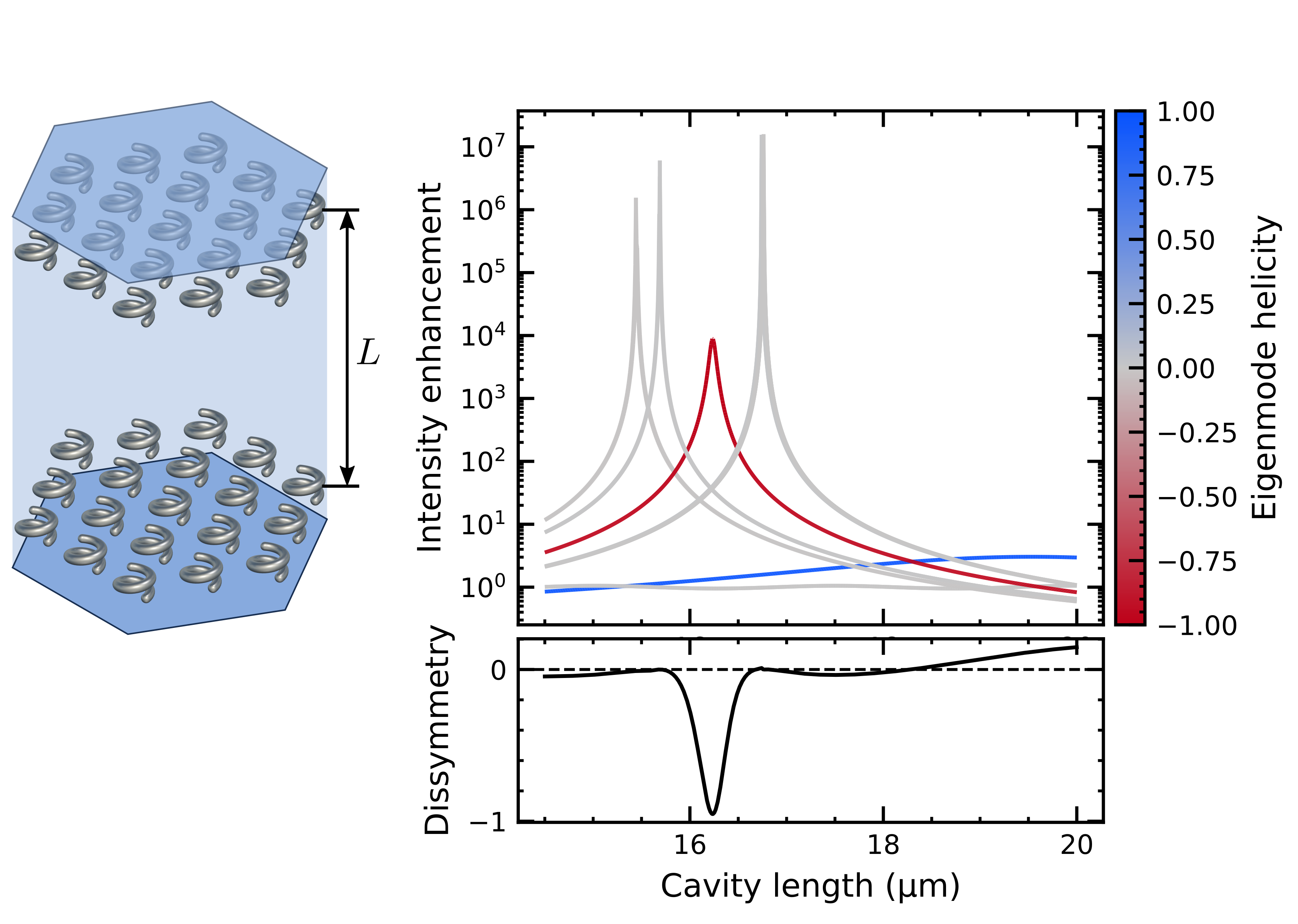}
	\caption{Internal intensity enhancement of cavity eigenmodes colored according to their helicity content (top), and total dissymmetry (bottom).
    The frequency is $\qty{44.21}{\THz}$, and eigenmodes that have a component with zero in-plane momentum are considered.
    Different eigenmodes are brought into and out of resonance by varying the cavity length $L$. 
    On resonance, the contrast in the intensity enhancement between the negative helicity and positive helicity modes is just shy of four orders of magnitude.
    The helicity of the dominant mode is $\approx \num{-0.988}$.
    Considering all these modes, at $L = \qty{16.23}{\um}$ the negative helicity dominates the cavity with a total dissymmetry equal to \num{-0.953}, so \qty{95}{\percent} of its absolute bound.}
    \label{fig:cavity-eigenmodes}
\end{figure}

Finally, we turn our attention to the chiral cavity, as depicted in \cref{fig:cavity-eigenmodes}, and the resonant modes, or eigenmodes, that it supports.
We leverage the S-matrix formalism, in which every S-matrix represents a two-dimensionally periodic system.
The S-matrix maps the expansion coefficients of incoming waves with well-defined wave vector and helicity, that is, monochromatic circularly polarized plane waves, to outgoing ones.
In other words, it holds the structure's reflection and transmission coefficients, including those related to higher-order diffraction.
The efficient computation of S-matrices for the system studied here, combining lattices and layers of homogeneous media, is made possible by the software package \pkg{treams}\footnote{\pkg{treams} version 0.4.5, published via \href{https://github.com/tfp-photonics/treams}{GitHub}.} \cite{beutel_treams_2024}.

The cavity is formed by three elements: The first two are the upper and lower mirror, $\Smat^{\text{upper}}$ and $\Smat^{\text{lower}}$, which account for the combined effects of the diffracting lattices of helices and the substrates, as explored above.
The third element is the cavity interior assumed to be filled with a material matching the permittivity $\permittivity_{\text{amb}}$. 
As the waves traverse the cavity interior from one mirror to the other they pick up a phase factor according to the out-of-plane component of their angular wave vector in the medium and the cavity length, $k_{z}L$.
The cavity interior is represented by another S-matrix, $\Smat^{\text{prop}}$, which holds these phase factors due to propagation on its diagonal.

Using these elements, a round trip inside the cavity is described by the term
\begin{equation} \label{eq:cavity-roundtrip}
    \Srt
    \coloneqq
    \Smat^{\text{lower}}_{\uparrow\downarrow} 
    \Smat^{\text{prop}}_{\downarrow\downarrow}
    \Smat^{\text{upper}}_{\downarrow\uparrow}
    \Smat^{\text{prop}}_{\uparrow\uparrow} \,.
\end{equation}
Here, the subscripts designate the common out-of-plane propagation directions of the outgoing and incoming waves in that order.
In this analysis, we consider plane waves at normal incidence to the mirror (zero in-plane momentum), as well as the related (six) first diffraction orders, for a total of seven different wave vectors.
As each wave vector is associated with two circularly polarized plane waves, one left-handed and one right-handed, the matrices in \cref{eq:cavity-roundtrip} are of dimension $14 \times 14$.
For computing the responses of the cavity mirrors, higher diffraction orders were considered as well.
However, these higher-order waves are evanescent in the cavity medium and decay quickly in the cavity interior.
Correspondingly, their influence on the resonant modes of the system can be neglected for the considered cavity lengths.

We can obtain the eigenmodes of the cavity from an eigendecomposition of $\Srt$.
Each eigenvector $\vec{v}_{i}$ of $\Srt$ defines a linear combination of the 14 basis waves that remains unchanged after a cavity round trip, except for a scaling factor $w_i$, which is the corresponding eigenvalue.
Furthermore, we can determine the internal field enhancement factors of each eigenmode inside the cavity, that is, the amplitude ratio and phase relation of each resonantly enhanced eigenmode as compared to its excitation from within the cavity.
It is found after considering the infinite series of consecutive reflections of the modes inside the cavity, ultimately leading to a geometric series in $\Srt$, yielding
\begin{equation}
\begin{split}
    \Sres
    &\coloneqq
    \left(
        \Id 
        - 
        \Srt
    \right)\inv \\
    &=
    \left(
        \Id 
        - 
        \Smat^{\text{lower}}_{\uparrow\downarrow} 
        \Smat^{\text{prop}}_{\downarrow\downarrow}
        \Smat^{\text{upper}}_{\downarrow\uparrow}
        \Smat^{\text{prop}}_{\uparrow\uparrow}
    \right)\inv \,.
\end{split}
\end{equation}
Convergence is guaranteed as the considered system is passive with non-vanishing losses.
The internal field enhancement of the cavity eigenmodes is given by the eigenvalues of $\Sres$.
Conveniently, the geometric series can also be carried out on the level of the eigenvalues $w_{i}$ of $\Srt$, from which we find the eigenvalues of $\Sres$ as $\frac{1}{1 - w_{i}}$.
From these, the internal intensity enhancement is simply found as the modulus squared, $\intensityenhancement_{i} \coloneqq \abs*{\frac{1}{1 - w_{i}}}^{2}$.

The internal intensity enhancement experienced by the cavity eigenmodes is plotted in \cref{fig:cavity-eigenmodes}. 
For the field inside the cavity to build up resonantly, the total phase that the mode accumulates in a cavity round trip needs to be an integer multiple of $2\ppi$. 
We vary the cavity length $L$ to tune different eigenmodes in and out of resonance. 

To study the helicity of the eigenmodes, we make use of the fact that the waves forming the basis for the S-matrices are of well-defined helicity.
The helicity of an eigenmode can then be evaluated from a weighted sum over its (normalized) eigenvector components,
\begin{equation}
    \Lambda(\vec{v}_{i})
    \coloneqq
    \sum_{n=1}^{14}
    \abs*{(v_{i})_{n}}^{2}
    \lambda_{n} \,,
\end{equation}
where $\lambda_{n} \in \{-1, +1\}$ is the helicity of the basis wave with index $n$.
As the eigenvector is normalized, $\Lambda(\vec{v}_{i})$ is bound between $-1$ and $1$, reaching the extreme values only for eigenmodes of pure helicity, that is, composed of exclusively right-handed or left-handed circularly polarized plane waves, respectively.
We evaluate the helicity of the eigenmodes and color the related intensity enhancement in \cref{fig:cavity-eigenmodes} correspondingly, individually for every considered cavity length.
The helicity of the eigenmodes changes only marginally when changing the cavity length, and, accordingly, the lines plotted in \cref{fig:cavity-eigenmodes} appear to be uniformly colored. 
Among the eigenmodes supported by the cavity, two stand out as being of almost pure helicity at $\abs*{\Lambda} \approx \num{0.988}$, one of each sign. 
However, while the negative helicity cavity mode reaches internal intensity enhancements of up to $\num{8.89e3}$, the positive helicity mode reaches a mere $\num{3.04}$, even when tuned on resonance at a cavity length of \qty{19.53}{\um}. 
Importantly, the resonance condition for both modes is not met simultaneously. 
Therefore, at the resonance of the negative helicity eigenmode at $L = \qty{16.23}{\um}$, there is a contrast of close to four orders of magnitude in the intensity enhancement between the negative helicity and positive helicity eigenmodes.

Finally, to estimate how strongly the considered eigenmode landscape inside the cavity is dominated by one helicity, we assume for simplicity that all eigenmodes are excited equally and compute the following dissymmetry measure
\begin{equation}
    \gamma \coloneqq \frac{\sum_{i}\Lambda(\vec{v}_{i})\intensityenhancement_{i}}{\sum_{i}\intensityenhancement_{i}} \,
\end{equation}
where $\intensityenhancement_{i}$ refers to the intensity enhancement of eigenmode $i$. 
On resonance at $L = \qty{16.23}{\um}$, it reaches ${\gamma \approx \num{-0.953}}$, so \qty{95}{\percent} of its absolute bound. 
The bottom panel of \cref{fig:cavity-eigenmodes} shows $\gamma$ as a function of $L$.

In conclusion, we have presented a chiral infrared cavity made from diffracting lattices of scatterers with high electromagnetic chirality. 
The unprecedented dissymmetry inside the cavity at the target frequency makes it a compelling candidate for enantio-selective applications in optical cavities.

\begin{acknowledgments}
L.R. acknowledges support by the Karlsruhe School of Optics \& Photonics (KSOP).
I.F.C. and C.R. acknowledge support by the Helmholtz Association via the Helmholtz program ``Materials Systems Engineering'' (MSE).
This work was partially funded by the Deutsche Forschungsgemeinschaft (DFG, German Research Foundation) -- Project-ID 258734477 -- SFB 1173.
We are grateful to the company JCMwave for their free provision of the FEM Maxwell solver \pkg{JCMsuite}. 
\end{acknowledgments}

\appendix

\section{Transition matrices from finite element method simulations}
\label{sec:tmatrix_fem}

The linear optical response of an isolated localized object can be represented as a \emph{transition matrix}, or \emph{T-matrix} \cite{waterman_matrix_1965, waterman_symmetry_1971, mishchenko_peter_2013}.
The T-matrix is typically formulated in the frequency domain, describing time-harmonic, monochromatic fields.
In the context of electromagnetic scattering off of the object, the T-matrix maps the coefficients of the incident field, expanded into a basis of regular \acp{VSW}, to the coefficients of the scattered field, itself expanded into singular (outgoing) \acp{VSW}.
Given an object's T-matrix, its scattering response to an arbitrary incident field (represented as a column vector of expansion coefficients) is readily computed as a basic matrix--vector product.
Furthermore, some quantities of interest, such as orientationally averaged extinction, scattering, and absorption cross-sections, or the em-chirality can be directly computed in terms of the T-matrix without considering a specific illumination \cite{mishchenko_tmatrix_1996, fernandez-corbaton_objects_2016}.

To obtain T-matrices of nontrivially shaped, that is, nonspherical, scatterers, we perform scattering simulations using the \ac{FEM} solver \pkg{JCMsuite}. 
The setup consists of the scatterer, as represented by a spatial distribution of optical material parameters localized around the origin, embedded in a finite computational domain made of a lossless, homogeneous, and isotropic ambient material.
The finite computational domain is padded with \acp{PML} to suppress backscattering of outgoing fields, thereby modeling the ambient material to be infinitely extended.

Following the definition of the T-matrix given above, we choose \acp{VSW} as the sources of illumination in the \ac{FEM} scattering simulations.
The resulting scattered field for a given illuminating \ac{VSW} (of unit amplitude) is expanded into \acp{VSW} using a built-in postprocessing procedure of \pkg{JCMsuite}.
The expansion coefficients exactly correspond to the entries of one column of the T-matrix.
Repeating this scheme using one \ac{VSW} at a time for the illumination, the T-matrix can be populated column by column.
In practice, the bases of regular and singular \acp{VSW} are truncated to include only \acp{VSW} of multipolar degree $\degree$ (positive integer total angular momentum number) up to some maximum degree $\degree_\text{max}$, including all multipolar orders $\order$ (integer angular momentum projection onto $z$-axis) with $\abs*{\order} \leq \degree$.
This leaves us with a finite (truncated) representation of the T-matrix.
The maximum multipolar degree required to accurately represent a scatterer's optical response generally increases with the scatterer's size relative to the wavelength of the illumination in the ambient medium.
To obtain spectrally resolved data, the sketched procedure is repeated frequency by frequency, potentially adjusting the simulation setup to account for dispersive material parameters.

We would like to refer the reader to Ref.~\onlinecite{asadova_tmatrix_2025}, where a comprehensive file format for saving T-matrix data adhering to the FAIR principles is specified.
T-matrices computed as part of the work presented here were stored accordingly.
Furthermore, Ref.~\onlinecite{asadova_tmatrix_2025} contains example scripts for the numerical computation of T-matrices, including some that implement the same \pkg{JCMsuite}-based procedure as described here.

\section{Optimizing electromagnetically chiral helices}
\label{sec:optimization}

We employ an optimization routine, similar to the one discussed in Ref.~\onlinecite{garcia-santiago_maximally_2022}, to obtain an object that exhibits a strong dissimilarity in its optical response to illumination of one helicity, or polarization handedness, versus the other.
Reference~\onlinecite{fernandez-corbaton_objects_2016} formulates a measure to quantify the difference in the helicity-dependent response, that is, the object's \emph{chiral} optical response.
This measure, the em-chirality, serves as the objective function for our optimization.

The optimization is performed on metallic (micro)helices, which have been shown to be able to reach high values of em-chirality before \cite{garcia-santiago_maximally_2022}.
The helix (implemented as a 3D primitive in \pkg{JCMsuite}) is centered at the origin and embedded in a lossless, isotropic, and homogeneous ambient material with relative permittivity $\permittivity_{\text{amb}} = \num{1.8912}$.
The illumination frequency is set to $\qty{44.21}{\tera\hertz}$.
The helix is assigned material parameters that model silver, from a quadratic interpolation of the data published in Ref.~\onlinecite{hagemann_optical_1975} evaluated at the illumination frequency. 

The helix is geometrically specified by four geometry parameters.
These parameters are subject to the later optimization.
Namely, these are the \emph{major radius}, $R$, the radius of the helical windings around the helix axis, the \emph{minor radius}, $r$, the radius of the circular cross-section of the wound up silver wire, the \emph{pitch}, $p$, the axial distance between consecutive windings, and the \emph{number of turns}, $n_\text{turns}$, or total winding number of the helix.
The wound-up silver wire that composes the helix further terminates in hemispherical end-caps.

We optimize the helix's geometry parameters with respect to its em-chirality, using a Bayesian optimization scheme as provided by \pkg{JCMsuite}'s optimization toolkit.
The objective function takes the four geometry parameters listed above as inputs and outputs the (normalized) em-chirality of the thereby specified helix.
As the em-chirality is defined in terms of the T-matrix, the first step in evaluating the objective function consists of computing the T-matrix of the specified helix as described in \cref{sec:tmatrix_fem}.
Then, the normalized em-chirality $\emchirality(\Tmat)$ is evaluated, using the method provided by \pkg{treams}.
Evaluating the objective function is computationally expensive, as it involves multiple full-wave simulations.
For this reason, an optimization technique that requires relatively fewer function calls, such as Bayesian optimization, is favorable \cite{schneider_benchmarking_2019}.
Finally, the geometry parameters that yield the highest em-chirality define the optimized helix used in the further process.

\bibliography{references}

\end{document}